\documentclass{preprint}
\usepackage{helvet}
\usepackage{courier}
\usepackage[T1]{fontenc}
\usepackage[a4paper]{geometry}
\usepackage[active]{srcltx}
\usepackage{color}
\usepackage{verbatim}
\usepackage{mathrsfs}
\usepackage{mathtools}
\usepackage{amsmath}
\usepackage{amsthm}
\usepackage{amssymb}
\usepackage{makeidx}
\makeindex
\usepackage[pdfusetitle,
 bookmarks=true,bookmarksnumbered=true,bookmarksopen=true,bookmarksopenlevel=2,
 breaklinks=false,pdfborder={0 0 1},backref=page,colorlinks=false]
 {hyperref}
\hypersetup{
 unicode=false, linkcolor=black, citecolor=black, urlcolor=blue, filecolor=blue, pdfpagelayout=OneColumn, pdfnewwindow=true, pdfstartview=XYZ, plainpages=false}

\makeatletter
\numberwithin{equation}{section}
\theoremstyle{plain}
\newtheorem{thm}{\protect\theoremname}[section]
\theoremstyle{plain}
\newtheorem{fact}[thm]{\protect\factname}
\theoremstyle{definition}
\newtheorem{defn}[thm]{\protect\definitionname}
\theoremstyle{plain}
\newtheorem{question}[thm]{\protect\questionname}
\theoremstyle{plain}
\newtheorem{ax}[thm]{\protect\axiomname}

\usepackage{color}

 \let\myTOC\tableofcontents
 \renewcommand\tableofcontents{%
   \pdfbookmark[1]{Contents}{}
   \myTOC
   \cleardoublepage
   \pagenumbering{arabic} }

\global\long\def\foreignlanguage#1#2{#2}%
\global\long\def\selectlanguage#1{}%

\allowdisplaybreaks

\makeatother

\providecommand{\axiomname}{Axiom}
\providecommand{\definitionname}{Definition}
\providecommand{\factname}{Fact}
\providecommand{\questionname}{Question}
\providecommand{\theoremname}{Theorem}

\begin{document}
\title{A note on the conceptual problems on the Unruh effect}
\author{Hideyasu Yamashita}
\institute{Division of Liberal Arts and Sciences, Aichi-Gakuin University\\
\email{yamasita@dpc.aichi-gakuin.ac.jp}}

\date{\today}

\maketitle
\newcommand{\dcolor}{\definecolor{note_fontcolor}{rgb}{0.1, 0.0, 0.8}}
\definecolor{HYnote_fontcolor}{rgb}{0.2, 0.0, 0.2}
\definecolor{hycolor}{rgb}{0.3, 0.0, 0.3}
\newcommand{\hyc}{\color{hycolor}}
\newenvironment{HYnote}
 {\textcolor{note_fontcolor}\bgroup\ignorespaces}
  {\ignorespacesafterend\egroup} 

\newenvironment{trivenv}
  {\bgroup\ignorespaces}
  {\ignorespacesafterend\egroup}

\newcommand{\displabel}[1]{}

\newcommand{\hidable}[3]{#2}
\newcommand{\hidea}[1]{{#1}}
\newcommand{\hideb}[1]{{#1}}
\newcommand{\hidec}[1]{{#1}}
\newcommand{\hidep}[1]{{#1}}
\renewcommand{\hidec}[1]{}
\renewcommand{\hidep}[1]{}

\newcommand{\thlab}[1]{{\tt [#1]}}

\newcommand{\black}{\color{black}}

\global\long\def\N{\mathbb{N}}%
\global\long\def\C{\mathbb{C}}%
\global\long\def\Z{\mathbb{Z}}%
 
\global\long\def\R{\mathbb{R}}%
 
\global\long\def\im{\mathrm{i}}%

\global\long\def\di{\partial}%
 
\global\long\def\d{{\rm d}}%

\global\long\def\ol#1{\overline{#1}}%
\global\long\def\ul#1{\underline{#1}}%
\global\long\def\ob#1{\overbrace{#1}}%

\global\long\def\ov#1{\overline{#1}}%

\global\long\def\then{\Rightarrow}%
 
\global\long\def\Then{\Longrightarrow}%

\global\long\def\N{\mathbb{N}}%
\global\long\def\C{\mathbb{C}}%
\global\long\def\Z{\mathbb{Z}}%
 
\global\long\def\R{\mathbb{R}}%
 
\global\long\def\im{\mathrm{i}}%

\global\long\def\di{\partial}%
 
\global\long\def\d{{\rm d}}%

\global\long\def\ol#1{\overline{#1}}%
\global\long\def\ul#1{\underline{#1}}%
\global\long\def\ob#1{\overbrace{#1}}%

\global\long\def\ov#1{\overline{#1}}%

\global\long\def\then{\Rightarrow}%
 
\global\long\def\Then{\Longrightarrow}%

\global\long\def\cA{\mathcal{A}}%
\global\long\def\cB{\mathcal{B}}%
 
\global\long\def\cC{\mathcal{C}}%
 
\global\long\def\cD{\mathcal{D}}%
\global\long\def\cE{\mathcal{E}}%
 
\global\long\def\cF{\mathcal{F}}%
 
\global\long\def\cG{{\cal G}}%
 
\global\long\def\cH{\mathcal{H}}%
 
\global\long\def\cI{\mathcal{I}}%
 
\global\long\def\cJ{\mathcal{J}}%
\global\long\def\cK{\mathcal{K}}%
 
\global\long\def\cL{\mathcal{L}}%
 
\global\long\def\cM{\mathcal{M}}%
 
\global\long\def\cN{\mathcal{N}}%
 
\global\long\def\cO{\mathcal{O}}%
 
\global\long\def\cP{\mathcal{P}}%
 
\global\long\def\cQ{\mathcal{Q}}%
 
\global\long\def\cR{\mathcal{R}}%
 
\global\long\def\cS{\mathcal{S}}%
 
\global\long\def\cT{\mathcal{T}}%
 
\global\long\def\cU{\mathcal{U}}%
 
\global\long\def\cV{\mathcal{V}}%
 
\global\long\def\cW{\mathcal{W}}%
\global\long\def\cX{\mathcal{X}}%
 
\global\long\def\cY{\mathcal{Y}}%
 
\global\long\def\cZ{\mathcal{Z}}%

\global\long\def\scA{\mathscr{A}}%
\global\long\def\scB{\mathscr{B}}%
\global\long\def\scC{\mathscr{C}}%
\global\long\def\scD{\mathscr{D}}%
 
\global\long\def\scE{\mathscr{E}}%
 
\global\long\def\scF{\mathscr{F}}%
 
\global\long\def\scG{\mathscr{G}}%
 
\global\long\def\scH{\mathscr{H}}%
 
\global\long\def\scI{\mathscr{I}}%
 
\global\long\def\scJ{\mathscr{J}}%
 
\global\long\def\scK{\mathscr{K}}%
 
\global\long\def\scL{\mathscr{L}}%
 
\global\long\def\scM{\mathscr{M}}%
 
\global\long\def\scN{\mathscr{N}}%
 
\global\long\def\scO{\mathscr{O}}%
 
\global\long\def\scP{\mathscr{P}}%
 
\global\long\def\scR{\mathscr{R}}%
\global\long\def\scS{\mathscr{S}}%
 
\global\long\def\scT{\mathscr{T}}%
 
\global\long\def\scU{\mathscr{U}}%
 
\global\long\def\scW{\mathscr{W}}%
\global\long\def\scZ{\mathscr{Z}}%

\global\long\def\bbA{\mathbb{A}}%
 
\global\long\def\bbB{\mathbb{B}}%
 
\global\long\def\bbD{\mathbb{D}}%
 
\global\long\def\bbE{\mathbb{E}}%
 
\global\long\def\bbF{\mathbb{F}}%
 
\global\long\def\bbG{\mathbb{G}}%
 
\global\long\def\bbI{\mathbb{I}}%
 
\global\long\def\bbJ{\mathbb{J}}%
 
\global\long\def\bbK{\mathbb{K}}%
 
\global\long\def\bbL{\mathbb{L}}%
 
\global\long\def\bbM{\mathbb{M}}%
 
\global\long\def\bbP{\mathbb{P}}%
 
\global\long\def\bbQ{\mathbb{Q}}%
 
\global\long\def\bbT{\mathbb{T}}%
 
\global\long\def\bbU{\mathbb{U}}%
 
\global\long\def\bbX{\mathbb{X}}%
 
\global\long\def\bbY{\mathbb{Y}}%
\global\long\def\bbW{\mathbb{W}}%

\global\long\def\bbOne{1\kern-0.7ex  1}%

\renewcommand{\bbOne}{\mathbbm{1}}

\global\long\def\bB{\mathbf{B}}%
 
\global\long\def\bG{\mathbf{G}}%
 
\global\long\def\bH{\mathbf{H}}%
\global\long\def\bS{\boldsymbol{S}}%
 
\global\long\def\bT{\mathbf{T}}%
 
\global\long\def\bX{\mathbf{X}}%
\global\long\def\bY{\mathbf{Y}}%
\global\long\def\bW{\mathbf{W}}%
 
\global\long\def\boT{\boldsymbol{T}}%

\global\long\def\fraka{\mathfrak{a}}%
 
\global\long\def\frakb{\mathfrak{b}}%
 
\global\long\def\frakc{\mathfrak{c}}%
 
\global\long\def\frake{\mathfrak{e}}%
 
\global\long\def\frakf{\mathfrak{f}}%
 
\global\long\def\fg{\mathfrak{g}}%
 
\global\long\def\frakh{\mathfrak{h}}%
 
\global\long\def\fraki{\mathfrak{i}}%
\global\long\def\frakk{\mathfrak{k}}%
 
\global\long\def\frakl{\mathfrak{l}}%
 
\global\long\def\frakm{\mathfrak{m}}%
 
\global\long\def\frakn{\mathfrak{n}}%
 
\global\long\def\frako{\mathfrak{o}}%
 
\global\long\def\frakp{\mathfrak{p}}%
 
\global\long\def\frakq{\mathfrak{q}}%
 
\global\long\def\fraks{\mathfrak{s}}%
 
\global\long\def\fs{\mathfrak{s}}%
 
\global\long\def\fraku{\mathfrak{u}}%
\global\long\def\frakz{\mathfrak{z}}%

\global\long\def\fA{\mathfrak{A}}%
 
\global\long\def\fB{\mathfrak{B}}%
 
\global\long\def\fC{\mathfrak{C}}%
 
\global\long\def\fD{\mathfrak{D}}%
 
\global\long\def\fF{\mathfrak{F}}%
 
\global\long\def\fG{\mathfrak{G}}%
 
\global\long\def\fK{\mathfrak{K}}%
 
\global\long\def\fL{\mathfrak{L}}%
 
\global\long\def\fM{\mathfrak{M}}%
 
\global\long\def\fP{\mathfrak{P}}%
 
\global\long\def\fR{\mathfrak{R}}%
 
\global\long\def\fS{\mathfrak{S}}%
\global\long\def\fT{\mathfrak{T}}%
 
\global\long\def\fU{\mathfrak{U}}%
 
\global\long\def\fX{\mathfrak{X}}%

\global\long\def\ssS{\mathsf{S}}%
\global\long\def\ssT{\mathsf{T}}%
 
\global\long\def\ssW{\mathsf{W}}%

\global\long\def\hM{\hat{M}}%

\global\long\def\rM{\mathrm{M}}%
\global\long\def\prj{\mathfrak{P}}%

{} 
\global\long\def\sy#1{{\color{blue}#1}}%

\global\long\def\magenta#1{{\color{magenta}#1}}%

\global\long\def\symb#1{#1}%

\global\long\def\emhrb#1{\text{{\color{red}{\huge {\bf #1}}}}}%

\newcommand{\symbi}[1]{\index{$ #1$}{\color{red}#1}} 

{} 
\global\long\def\SYM#1#2{#1}%

\renewcommand{\SYM}[2]{\symb{#1}}

\newcommand{\usuji}{\color[rgb]{0.7,0.4,0.4}} \newcommand{\usu}{\color[rgb]{0.5,0.2,0.1}}
\newenvironment{Usuji} {\begin{trivlist}   \item \usuji }  {\end{trivlist}}
\newenvironment{Usu} {\begin{trivlist}   \item \usu }  {\end{trivlist}} 

\newcommand{\term}[1]{\textcolor[rgb]{0, 0, 1}{\bf #1}}
\newcommand{\termi}[1]{{\bf #1}}

\global\long\def\supp{{\rm supp}}%
\global\long\def\dom{\mathrm{dom}}%
\global\long\def\ran{\mathrm{ran}}%
 
\global\long\def\leng{\text{{\rm leng}}}%
 
\global\long\def\diam{\text{{\rm diam}}}%
 
\global\long\def\Leb{\text{{\rm Leb}}}%
 
\global\long\def\meas{\text{{\rm meas}}}%
\global\long\def\sgn{{\rm sgn}}%
 
\global\long\def\Tr{{\rm Tr}}%
 
\global\long\def\tr{\mathrm{tr}}%
 
\global\long\def\spec{{\rm spec}}%
 
\global\long\def\Ker{{\rm Ker}}%
 
\global\long\def\Lip{{\rm Lip}}%
 
\global\long\def\Id{{\rm Id}}%
 
\global\long\def\id{{\rm id}}%

\global\long\def\ex{{\rm ex}}%
 
\global\long\def\Pow{\mathsf{P}}%
 
\global\long\def\Hom{\mathrm{Hom}}%
 
\global\long\def\div{\mathrm{div}}%
 
\global\long\def\grad{\mathrm{grad}}%
 
\global\long\def\Lie{{\rm Lie}}%
 
\global\long\def\End{{\rm End}}%
 
\global\long\def\Ad{{\rm Ad}}%

\newcommand{\slim}{\mathop{\mbox{s-lim}}} %

\newcommand{\wlim}{\mathop{\mbox{w-lim}}}

\newcommand{\limsub}{\mathop{\mbox{\rm lim-sub}}}

\global\long\def\bboxplus{\boxplus}%

\renewcommand{\bboxplus}{\mathop{\raisebox{-0.8ex}{\text{\begin{trivenv}\LARGE{}$\boxplus$\end{trivenv}}}}}

\global\long\def\shuff{\sqcup\kern-0.3ex  \sqcup}%

\renewcommand{\shuff}{\shuffle}

\global\long\def\upha{\upharpoonright}%

\global\long\def\ket#1{|#1\rangle}%
 
\global\long\def\bra#1{\langle#1|}%

{} 
\global\long\def\lll{\vert\kern-0.25ex  \vert\kern-0.25ex  \vert}%
 \renewcommand{\lll}{{\vert\kern-0.25ex  \vert\kern-0.25ex  \vert}}

\global\long\def\biglll{\big\vert\kern-0.25ex  \big\vert\kern-0.25ex  \big\vert\kern-0.25ex  }%
 
\global\long\def\Biglll{\Big\vert\kern-0.25ex  \Big\vert\kern-0.25ex  \Big\vert}%

\newcommand{\iiia}[1]{{\left\vert\kern-0.25ex\left\vert\kern-0.25ex\left\vert #1
  \right\vert\kern-0.25ex\right\vert\kern-0.25ex\right\vert}}

\global\long\def\iii#1{\iiia{#1}}%

\global\long\def\Upa{\Uparrow}%
 
\global\long\def\Nor{\Uparrow}%

\newcommand{\vertt}{\kern-0.6ex\vert}
\renewcommand{\Nor}{[\kern-0.16ex ]}

\global\long\def\Prob{\mathbb{P}}%
\global\long\def\Var{\mathrm{Var}}%
\global\long\def\Cov{\mathrm{Cov}}%
\global\long\def\Ex{\mathbb{E}}%
{} 
\global\long\def\Ae{{\rm a.e.}}%
 
\global\long\def\samples{\bOm}%


\global\long\def\bOne{{\bf 1}}%

\global\long\def\Ten{\bullet}%
{} %

\global\long\def\TT{\intercal}%
 \renewcommand{\TT}{\mathsf{T}}

\global\long\def\trit{\vartriangle\!\! t}%

\global\long\def\p{\mathbf{p}}%
 
\global\long\def\q{\mathbf{q}}%
 
\global\long\def\bA{\mathbf{A}}%
 
\global\long\def\x{\mathbf{x}}%
 
\global\long\def\y{\mathbf{y}}%

\global\long\def\WICK#1{{\rm w}\{#1\}}%
 \renewcommand{\WICK}[1]{{:}#1{:}}

\global\long\def\ssD{\boldsymbol{D}}%
 \renewcommand{\ssD}{\mathsf{D}}

\global\long\def\ssK{\boldsymbol{K}}%
 \renewcommand{\ssK}{\mathsf{K}}

\global\long\def\ssH{\boldsymbol{H}}%
 \renewcommand{\ssH}{\mathsf{H}}

\global\long\def\ssN{\boldsymbol{N}}%
 \renewcommand{\ssN}{\mathsf{N}}

\global\long\def\ssR{\boldsymbol{R}}%
 \renewcommand{\ssR}{\mathsf{R}}

\global\long\def\ssP{\boldsymbol{P}}%
 \renewcommand{\ssP}{\mathsf{P}}

\global\long\def\bQ{\mathbf{Q}}%
 
\global\long\def\rF{\mathrm{F}}%
 
\global\long\def\bM{\mathbf{M}}%
 
\global\long\def\rE{\mathrm{E}}%
 
\global\long\def\bV{\mathbf{V}}%

\global\long\def\bE{\mathbf{E}}%

\global\long\def\xsf{\mathsf{x}}%
 
\global\long\def\tsf{\mathsf{t}}%

\global\long\def\xtt{\mathtt{x}}%
 
\global\long\def\ttt{\mathtt{t}}%

\global\long\def\xtt{\underline{x}}%
 
\global\long\def\ttt{\underline{t}}%

\global\long\def\Temp{\mathsf{T}}%

\newcommand{\amp}{\,\&\,\ignorespaces}

\def\factname{Claim}

\def\foreignlanguage#1#2{#2}


\let\ruleorig=\rule
\renewcommand{\rule}{\noindent\ruleorig}

\global\long\def\labelenumi{(\arabic{enumi})}%

\begin{abstract}
This brief note is written with a somewhat similar purpose to Earman's
2011 paper on the conceptual problems on the Unruh effect. However,
we confine ourselves to Sewell's modular approach to the Unruh effect,
which is based on the theorems of Tomita-Takesaki and Bisognano\textendash Wichmann.
This approach is rigorous, and has an advantage of being model-independent.
However, we will see that a number of conceptual problems remain unsolved
on this approach.
\end{abstract}

\section{Introduction}

\label{sec:Introduction}

This brief note is written with a somewhat similar purpose to Earman's
2011 paper \cite{Ear2011} on the conceptual problems on the Unruh
effect. Since Earman is not only a philosopher of science, but also
a historian of science, Ref.~\cite{Ear2011} also serves very well
as a review article on the history of the study of Unruh effect, with
over 100 references. I refer to another good review article due to
Crispino, Higuchi and Matsas \cite{CHM2007}. Although I hope that
this note also serves as a review article, the content of this note
is far more biased than these papers, with much less references. 

We will begin with the following very rough statement of the \termi{Unruh effect}
\cite{Unr1976}.%

\begin{fact}
\label{claim:Unruh}In a Minkowski spacetime, a uniformly accelerating
observer in the vacuum ``feels'' the heat at the temperature proportional
to the acceleration. This temperature is called the \termi{Unruh temperature},
and given by
\begin{equation}
\Temp_{{\rm U}}=\frac{\hbar a}{2\pi ck_{{\rm B}}},\label{eq:UnruhTemp}
\end{equation}
where $a$ is the acceleration, $c$ is the speed of light, and $k_{{\rm B}}$
is the Boltzmann constant.
\end{fact}

I emphasize that it is quite difficult to give a precise definition
of the Unruh effect. %
Here I presupposed that a good definition of a physical effect should
make clear the empirical/operational meaning of the effect, i.e.,
that the definition should specify the experimental procedures to
verify/falsify the effect. However, there appears to be no common
understanding on the empirical meaning of the Unruh effect. In fact,
there are many papers proposing the experimental procedures to verify
the Unruh effect \cite[Sec.4]{CHM2007}, but some other authors criticize
some of them, and some still other authors suggest that such an experimental
procedure does not exist.\footnote{See e.g., \cite{BS2013,BV2015}. However their empirical consequences
are not very clear, and \cite{BS2013,BV2015} might actually intend
to say that there are procedures to falsify the Unruh effect.} In any case, there seems to be a vicious circle where the absence
of the precise definition prevents one from judging the validity of
such experimental proposals, and the absence of the reliable experimental
procedures prevents one from defining the Unruh effect.

Sewell%
{} \cite{Sew1982} gave a rigorous proof of the Unruh effect for the
first time. Another rigorous proof is found in De Bi\`{e}vre\amp
Merkli%
{} \cite{DBM2006}. In that sense, we may say that an ``existence''
of the Unruh effect is already established. However, as mentioned
above, the empirical meaning of the mathematical equation (\ref{eq:UnruhTemp})
is far from clear, and hence a mathematical proof of (\ref{eq:UnruhTemp})
does not immediately lead to the ``physical/empirical existence''
of the effect. In other words, a mathematically rigorous definition
of the Unruh effect is not necessarily a physically/empirically ``good''
definition of it.

Sometimes the Unruh effect is said to be a theoretical consequence
of quantum field theory in curved spacetime (QFTCS) (e.g., \cite{Wal1994}).
This is true, but it should be noted that usually the Unruh effect
is described on a subspace of a Minkowski spacetime, called the Rindler
wedge. In fact, although the above description of the Unruh effect
refers to the vacuum, we cannot define the vacuum in a generic curved
spacetime.

Claim \ref{claim:Unruh} refers to what an accelerated observer sees
(or feels). It is well known that Einstein employed the thought experiments
of this type, which led him to general relativity. If Claim \ref{claim:Unruh}
is a logical consequence of a theory, which theory can lead to that
consequence? Many authors, including Unruh himself \cite{Unr1976},
primarily employ QFT in Minkowski spacetime (QFTM), and sometimes
QFT in curved spacetime (QFTCS) to examine the Unruh effect. However,
probably we need the full theory of general relativity to obtain the
firm logical consequences on what an accelerated observer sees. Thus
we possibly need both QFT and general relativity, and perhaps also
quantum gravity, to obtain Claim \ref{claim:Unruh} with a firm ground.%
{} However, the theory of quantum gravity is unlikely to be completed
in the near future. Thus Claim \ref{claim:Unruh} seems logically
too strong, given the theories currently available to us.

Next we will consider the restatement of the Unruh effect in a logically
weaker, more moderate form. The restatement of the following form
is due to Unruh himself \cite{Unr1976} and others (see \cite[Sec.7]{Ear2011}):
\begin{fact}
\label{claim:Unruh2-object}In a Minkowski spacetime, a uniformly
accelerating object in the vacuum is eventually heated to the Unruh
temperature $\Temp_{{\rm U}}=\hbar a/(2\pi ck_{{\rm B}})$.
\end{fact}

Note that Unruh \cite{Unr1976} and many other authors use the term
``detector'' instead of ``object''. In Claim \ref{claim:Unruh2-object}
I used ``object'' since the usage of the term ``detector'' in
the literature on the Unruh effect is somewhat unusual in that it
does not refer to an experimental apparatus such as a thermometer
in the usual sense, but to a simple microscopic object, such as a
single particle. Also note that the less familiar term ``thermalized''
is often used instead of ``heated'' in the literature.

Claim \ref{claim:Unruh2-object} avoids referring to any observer,
accelerating or not. However there is a major question on Claim \ref{claim:Unruh2-object},
concerning the very concept of temperature: If an object is microscopic,
is it meaningful to speak about its thermodynamic properties? Of course,
Claim \ref{claim:Unruh2-object} is less problematic when the object
is macroscopic. However the papers which examine Claim \ref{claim:Unruh2-object}
with macroscopic objects appear to be rare. Probably this is because
of the difficulty of the simultaneously special-relativistic and quantum-mechanical
treatment of macroscopic objects.

Although it is not quite evident that we are permitted to call a quantum
ideal (free) gas a ``macroscopic object'', here let us assume that
the there is no major problematic points on the concept of the temperature
of a quantum ideal gas. (We already know that the concept of temperature
of a quantum ideal gas can be given a mathematically rigorous ground
by the theory of quantum Gibbs states, and more generally that of
KMS states \cite{BR97}. However note that the mathematical rigor
does not necessarily guarantee the conceptual clarity. For example,
the temperature is defined only if a physical object is in a KMS state,
but how can we observe whether the object is in a KMS state or not?
Even if the object is in a KMS state, the empirical/operational meaning
of temperature becomes obscurer when the density of the gas is very
low.)

A usual (free or non-free) QFTM, which is represented on a Hilbert
space $\cH$, and has a unique vacuum in $\cH$, such as a QFTM satisfying
the Wightman axioms, admits no KMS states (of positive temperature)
as density operators on $\cH$. However, in a generalized sense, we
can consider the states of positive temperature in QFTM, or more generally
in QFT in curved space time (see \cite{HW2015} for a rigorous formulation
of it). Sometimes such a QFT is called a \termi{thermal QFT}. Note
that roughly speaking, the theory of quantum ideal gas is considered
to be equivalent to the thermal (quasi-)free QFT. Hence we can speak
about the KMS states and the temperature of a quasi-free field, and
probably also about those of a non-free (interacting) QFT. Of course,
currently we have no rigorous construction of an interacting QFT in
$(3+1)$-dimensional Minkowski spacetime, but there are some rigorous
constructions of interacting thermal QFT in $(1+1)$-dimensional Minkowski
spacetime \cite{GJ2005,GJ2005b}.

The concept of the temperature of a quantum field seems more transparent
than that of a microscopic object. Then consider the following version
of the statement of the Unruh effect:
\begin{fact}
\label{claim:Unruh-thermal} For an observer having a uniform acceleration
$a>0$ in a Minkowski spacetime, the Minkowski vacuum becomes a thermal
state at the Unruh temperature $\Temp_{{\rm U}}=\hbar a/(2\pi ck_{{\rm B}})$.
\end{fact}

In most cases we refer to the temperature of an object only when it
is in a KMS (or Gibbs) state. Hence we consider the term ``thermal
state'' to be synonymous with ``KMS state'' in this article, while
some generalized notions of thermal states other than KMS states have
been proposed in the literature (e.g., \cite{BOR2002}). I used the
term ``thermal state'' in Claim \ref{claim:Unruh-thermal} simply
because it is intuitively easier to grasp.

Since Claim \ref{claim:Unruh-thermal} refers to a uniformly accelerating
observer, it has a conceptual problem similar to that of Claim \ref{claim:Unruh}.
Furthermore, the meaning of ``becomes'' is dubious; Usually a transformation
of quantum states is represented by a unitary operator, but there
is no unitary transformation where the Minkowski vacuum state transforms
into a KMS state.

\rule[0.5ex]{1\columnwidth}{1pt}

{} Now we describe the KMS states slightly more precisely. (More rigorous
definition will be given later.) Let $\hbar=c=k_{{\rm B}}=1$ in the
following. Let $\cX$ be a (possibly curved) spacetime with the Lorentzian
metric $g$, and assume that $\cX$ is stationary, that is, that there
exists a complete time-like Killing vector field $\xi$ on $\cX$.
Interpret $\xi$ as the generator of a time flow on $\cX$, then $(\cX,g,\xi)$
is seen as a spacetime given a fixed time flow.%
{} The notion of the KMS state at the temperature $\Temp$ (or at the
inverse temperature $\beta:=1/\Temp$%
) is defined for each $(\cX,g,\xi)$. Especially, the absolute temperature
$\Temp$ is not actually ``absolute'' but ``relative'' to $\xi$
(\cite{HW2015}, see also \cite[Definition 9.11]{Ger2019}). If $(\cX,g)$
is fixed, let us say that a state $\omega$ is a \termi{$(\xi,\Temp)$-KMS state}
if $\omega$ is a KMS state at the temperature $\Temp>0$ with respect
to $\xi$.

Let $\{(t,x,y,z)\in\R^{4}\}$ be a coordinate system of a Minkowski
spacetime $\cM_{4}$. Consider an observer $O_{a}$ having constant
proper acceleration $a>0$ in $x$-direction. Let $(t,x,y,z)=(t(\tau),x(\tau),y(\tau),z(\tau))$
be the worldline of $O_{a}$, parametrized by the proper time $\tau$
of $O_{a}$. Up to translation, this is expressed as
\[
x=\frac{1}{a}\cosh(a\tau),\qquad t=\frac{1}{a}\sinh(a\tau),\qquad y=z=0,
\]
and implicitly, the world line of $O_{a}$ is
\[
x^{2}-t^{2}=\frac{1}{a^{2}},\qquad y=z=0.
\]
Consider the ``natural'' coordinate system $(T,X,Y,Z)$ for the
observer $O_{a}$, or ``the natural reference frame of the world
viewed from $O_{a}$''. This is given as follows, as will be discussed
later.
\[
x=X\cosh(aT),\quad t=X\sinh(aT),\quad y=Y,\quad z=Z,\qquad X>0
\]
and explicitly
\[
T=\frac{1}{a}{\rm arctanh}\frac{t}{x}=\frac{1}{2a}\ln\frac{x-t}{x+t},\qquad X=\sqrt{x^{2}-t^{2}},\quad Y=y,\quad Z=z.
\]
This coordinate system does not cover all of $\cM$; it only covers
the (right) \termi{Rindler wedge} $\cR=\cR_{{\rm r}}\subset\cM$
defined by
\[
\cR:=\{(t,x,y,z)\in\R^{4}:\ x>|t|\}.
\]
The Rindler wedge $\cR$ is a ``sub-spacetime'' of $\cM$, that
is, $\cR$ itself is a spacetime whose Lorentzian metric is the restriction
of that of $\cM$. Furthermore, $\cR$ is given the complete time-like
Killing vector field $\xi=\di/\di T$, and hence is a stationary spacetime.
Since $\xi$ is proportional to $a$, we will write $a\xi$ for $\xi$
in the following, letting $\xi$ denote the Killing vector field only
for $a=1$. Thus we can consider the $(a\xi,\Temp)$-KMS states on
$\cR$ for each $a>0$.

Roughly, the following statement is the formulation of the Unruh effect
due to Sewell \cite{Sew1982}, based on Bisognano \& Wichmann\cite{BW1975}.
\begin{fact}
\label{claim:Unruh-xi}Let $\xi$ be the time-like Killing vector
field on the Rindler wedge $\cR$ defined above. Let $\omega_{\cR}$
denote the state in the Rindler wedge $\cR$, given by restricting
the Minkowski vacuum $\omega$ onto $\cR$. Then $\omega_{\cR}$ is
a $(a\xi,\Temp_{{\rm U}})$-KMS state for all $a>0$ ($\Temp_{{\rm U}}:=a/(2\pi)$).
\end{fact}

Claim \ref{claim:Unruh-xi} itself does not refer to any observer,
and does not contain the dubious word ``becomes''. Hence it seems
free from the problematic points of Claim \ref{claim:Unruh-thermal}.
However note that Claim \ref{claim:Unruh-xi} is the most abstract
of Claims \ref{claim:Unruh}\textendash \ref{claim:Unruh-xi}; It
is rather closer to a purely mathematical statement, and hence it
is far from clear that Claim \ref{claim:Unruh-xi} is a physical/empirical
statement, which can be verified/falsified experimentally. Hence Claim
\ref{claim:Unruh-xi} itself might be too abstract to be called a
statement of the Unruh effect.

\rule[0.5ex]{1\columnwidth}{1pt}

Seemingly Claim \ref{claim:Unruh2-object} is conceptually the clearest
of Claims \ref{claim:Unruh}\textendash \ref{claim:Unruh-xi}, if
the accelerating object is macroscopic. However, the mainstream of
the study on Claim \ref{claim:Unruh2-object} including Unruh himself
\cite{Unr1976}, which is called the \termi{detector approach} in
\cite{Ear2011}, has been concentrated on the cases where the object
is microscopic, as mentioned above. In what follows we will examine
Claim \ref{claim:Unruh-xi} mainly.

\section{Temperature and time}

Before Claim \ref{claim:Unruh-xi}, the notion of temperature was
described to be relative to the time flow on a stationary spacetime,
such as the Rindler wedge. Given such a ``relativistic'' notion
of temperature, the observer-dependence of temperature seems somewhat
even tautological. However, this represents a radical shift in thinking
on the notion of temperature; Since conventional thermodynamics does
not appear to contain such understanding of temperature, we are required
to reconsider what is temperature. Although this section might seem
to digress somewhat from the main topic on the Unruh effect, I think
that this reconsideration is inevitable to understand the conceptual
aspects of the Unruh effect.

\subsection{KMS state and the Tomita\textendash Takesaki theorem}

Recall some basic definitions on KMS states and the Tomita\textendash Takesaki
theorem.

Let $\cA$ be a von Neumann algebra, and $\omega:\cA\to\C$ be a state
on $\cA$. That is, $\omega:\cA\to\C$ is a linear functional such
that $\omega(A^{*}A)\ge0$ and $\omega({\bf 1})=1$. Let $\{\sigma_{t}|t\in\R\}$
be a one-parameter group of automorphisms on $\cA$. Let $\beta>0$,
called the inverse temperature. Roughly speaking, $\omega$ is called
a {$(\sigma,\beta)$-KMS state} if $\omega(A\sigma_{i\beta}(B))=\omega(BA)$
for $A,B\in\cA$. Here $\sigma_{i\beta}$ is defined to be a suitable
analytic continuation of $\sigma_{t}$. Precisely,
\begin{defn}
$\omega$ is called a \termi{$(\sigma,\beta)$-KMS state} on $\cA$
if for any $A,B\in\cA$, there exists a complex function $f_{A,B}(z)$
defined on the strip $0\le{\rm Im}\,z\le\beta$ such that $f_{A,B}$
is analytic on $0<{\rm Im}\,z<\beta$, and%
\[
f_{A,B}(t)=\omega(B\sigma_{t}(A)),\qquad f_{A,B}(t+i\beta)=\omega(\sigma_{t}(A)B),\qquad\forall t\in\R.
\]
\end{defn}

In this note we will write ``\termi{$(\sigma,\Temp)$-KMS state}''
for ``$(\sigma,\beta)$-KMS state'' ($\beta=1/\Temp$ with $k_{{\rm B}}=1$)
in the following. Physically, a $(\sigma,\Temp)$-KMS state is interpreted
as a thermal equilibrium state where the temperature is $\Temp$ with
respect to the time evolution $\sigma$. Let $a>0$, and $\sigma_{t}':=\sigma_{at}$.
Then we see that $\omega$ is a $(\sigma,\Temp)$-KMS state if and
only if $\omega$ is a $(\sigma',a\Temp)$-KMS state. That is, if
the time scale (the ``speed of time'') is multiplied by $a$, the
temperature also becomes $a$ times greater.

A state $\omega$ on $\cA$ is \termi{normal} if it is completely
additive $\omega(\sum_{i}p_{i})=\sum_{i}\omega(p_{i})$ for any family
of pairwise orthogonal projections $\{p_{i}\}$ in $\cA$. This is
equivalent to that $\omega(A)=\Tr\rho A$ for some density operator
on $\cH$. A state $\omega$ is \termi{faithful} if $A\ge0,A\neq0\then\omega(A)>0$,
equivalently $\omega(A^{*}A)=0\then A=0$.
\begin{thm}
[Tomita--Takesaki theorem]\label{thm:TT}If $\omega$ is a faithful
normal state on $\cA$, there exists a unique one-parameter group
of automorphisms $\{\sigma_{t}|t\in\R\}$ on $\cA$ such that $\omega$
is a $(\sigma,1)$-KMS state. This $\sigma$ is called the \termi{group of modular automorphisms}
determined by $\omega$.
\end{thm}

\subsection{Connes\textendash Rovelli modular temperature hypothesis}

The Tomita\textendash Takesaki theorem itself is a purely mathematical
theorem. However it is reasonable to expect that it admits some physical
interpretations (e.g., \cite{CR1994}). First, the Tomita\textendash Takesaki
theorem is intuitively restated as follows:
\begin{itemize}
\item Fix $\Temp=1/\beta>0$. For any faithful normal state $\omega$, there
exists a unique time evolution $\sigma$ such that $\omega$ is a
thermal state at temperature $\Temp$ with respect to $\sigma$. The
speed of the evolution $\sigma$ is proportional to $\Temp$.
\end{itemize}
Note that the condition for a state $\omega$ to be normal is similar
to the complete additivity of the (classical) probability measure.
In fact, a probability distribution can be seen as a normal state
on an abelian von Neumann algebra. Thus we may expect that almost
all ``physical'' states are normal. Furthermore, note that a probability
distribution on $\R^{n}$ which has a non-vanishing density function
$\rho:\R^{n}\to(0,\infty)$, such as a Gaussian distribution, is seen
as a faithful normal state on the abelian von Neumann algebra $L^{\infty}(\R^{n})$.
Thus probably we may consider the faithful normal states to be rather
``generic'' or ``commonplace'' states in physics. Hence the Tomita\textendash Takesaki
theorem appears to have a surprising physical interpretation: For
any fixed $\Temp>0$, almost every ``physical state'' is a thermal
equilibrium state at temperature $\Temp$ with respect to a unique
time evolution, very roughly speaking.

Connes and Rovelli \cite{CR1994} (see also \cite[Sec.3.4]{Rov2004})
argued that this interpretation \emph{defines} a notion of time, called
the \termi{thermal time}, in quantum gravity and QFTCS. In fact,
the question ``what is time?'' already arises in classical general
relativity:
\begin{quote}
In a general covariant theory there is no preferred time flow, and
the dynamics of the theory cannot be formulated in terms of an evolution
in a single external time parameter. 

{[}...{]}
\end{quote}
\begin{quote}
One can still recover weaker notions of physical time: in GR, for
instance, on any given solution of the Einstein equations one can
distinguish timelike from spacelike directions and define proper time
along timelike world lines. {[}...{]}

Furthermore, this weaker notion of time is lost as soon as one tries
to include either thermodynamics or quantum mechanics into the physical
picture, because, in the presence of thermal or quantum \textquotedblleft superpositions\textquotedblright{}
of geometries, the spacetime causal structure is lost. This embarrassing
situation of not knowing \textquotedblleft what is time\textquotedblright{}
in the context of quantum gravity has generated the debated issue
of time of quantum gravity. \cite[Sec.1]{CR1994}
\end{quote}
Thus Connes and Rovelli \cite{CR1994} are concerned with ``what
is time'', rather than ``what is temperature''. On the other hand,
I am more concerned with the latter question, in this note. However,
now it is clear that these two questions are related to each other.
The most rough statement of this relation is:
\begin{fact}
\label{claim:temp=00003Dtime}the temperature is equal (or proportional)
to the ``speed of time (evolution)''.
\end{fact}

Roughly speaking, Claim \ref{claim:temp=00003Dtime} is what is called
the \termi{restricted modular temperature hypothesis} (RMTH) by Earman
\cite{Ear2011}.
For an observer $O$, the natural ``speed of time'' is determined
by the proper time of $O$. In other words, the natural ``speed of
time'' of w.r.t.~a worldline is determined by its proper time.

However some questions arise immediately:
\begin{question}
\label{que:propertime-extended}Generally, the proper time w.r.t.~a
worldline $C$ is meaningful only at each point of $C$ (or in a small
neighborhood of $C$). However, the thermodynamic notions, such as
temperature, are usually considered to be meaningful only in a sufficiently
extended spacetime region. How can we resolve this discrepancy? 
\end{question}

In fact, the KMS states in QFTCS is defined on the \emph{whole} spacetime
$M$, relatively to the time flow (the time-like Killing vector field)
on $M$, as mentioned above (\cite{HW2015}, \cite[Definition 9.11]{Ger2019}).
\begin{question}
\label{que:time-temp-empiri}What is the operational/empirical meaning
of Claim \ref{claim:temp=00003Dtime}? Claim \ref{claim:temp=00003Dtime}
seems to imply that the measurement of temperature is reduced to some
measurements of time. But to which measurements of time?
\end{question}

We already know that the technologies of atomic clocks and optical
lattice clocks enabled extremely precise time measurement. Hence if
temperature measurement can be reduced to time measurement, that would
be good news for us. However, I cannot imagine how to realize that.
Note that we have the time-energy uncertainty relation, while the
temperature of an object is roughly proportional to the internal energy
of it.

Currently I have no answer to Question \ref{que:time-temp-empiri},
and I will examine Question \ref{que:propertime-extended} in the
following.

\section{Rindler coordinate}

\label{sec:Rindler-coordinate}

Recall the Rindler coordinate mentioned in Sec.~\ref{sec:Introduction}.
Let $\{(t,x,y,z)\in\R^{4}\}$ be a coordinate system of a $(1+3)$-dimensional
Minkowski spacetime $\cM_{4}$. Let $O_{a}$ be an observer having
constant proper acceleration $a>0$ in $x$-direction, whose worldline
is parametrized by the proper time $\tau$ as follows.
\[
x=\frac{1}{a}\cosh(a\tau),\qquad t=\frac{1}{a}\sinh(a\tau),\qquad y=z=0.
\]
The \termi{Rindler coordinate system} $(T,X,Y,Z)$ on the \termi{Rindler wedge}
$\cR:=\{(t,x,y,z)\in\R^{4}:\ x>|t|\}$ is given by 
\begin{equation}
x=X\cosh(aT),\quad t=X\sinh(aT),\quad y=Y,\quad z=Z,\qquad X>0.\label{eq:RindlerCoord}
\end{equation}
I will give an argument that the Rindler wedge is the ``whole spacetime
viewed from $O_{a}$'', and the Rindler coordinate system is the
natural reference frame for $O_{a}$. This can be an answer to Question
\ref{que:propertime-extended}, because the Rindler coordinate system
defines a time flow on the whole Rindler spacetime (i.e., the Rindler
wedge viewed as a stationary spacetime), not only on a vicinity of
the worldline of $O_{a}$.

However, the meaning of \textquotedblleft natural coordinate system
for an observer (usually a person)\textquotedblright{} is too vague,
so let us consider a more concrete setting as follows. For simplicity,
we assume the laboratory containing the observer (and some observation
apparatus) is cube-shaped and rigid (thus its shape and size do not
change over time). We wish to consider the case where the laboratory
has constant acceleration, but we assume the laboratory is not rotating.
(The presence or absence of rotation can be detected by a gyroscope
installed in the laboratory.) Let the center of this cube be O. Using
this as the origin, a natural orthogonal coordinate system with the
length of one side of the cube as the unit seems conceivable.

However, this assumption presents some problems already at the stage
of special relativity, even without considering general relativity
or curved spacetime, because in special relativity, rigid bodies cannot
exist as physical objects. (Since the speed of sound propagating through
a rigid body is infinite, information could be transported faster
than light.)

If we consider this cube not as a physical object but as a more idealized
figure (a mathematical object), the contradiction with relativistic
causality does not arise. However, in this case, the meaning of calling
this cube a \textquotedblleft laboratory\textquotedblright{} becomes
diminished. If the size or shape of the laboratory has observational
significance, this is a problem that cannot be ignored. But from this
point onward, we will adopt the policy of considering only situations
where this is not the case (that is, we consider this cube itself
not to function as an observational apparatus, but merely as an \textquotedblleft virtual
frame\textquotedblright{} surrounding the observer).

For now, let us accept this idealization (regarding the laboratory
as an idealized rigid body) and assume that a \textquotedblleft natural
coordinate system for the observer,\textquotedblright{} in the sense
described above, exists at least in the vicinity of the laboratory.\footnote{This setting is similar to that of Buchholz\amp Verch \cite{BV2015},
while they do not mention the problem on rigid body in special relativity.}

\rule[0.5ex]{1\columnwidth}{1pt}

The Rindler wedge $\cR$ is a spacetime (Lorentz manifold) with the
time flow determined by the time axis $T$. The Minkowski metric is
expressed on $\cR$ by
\[
ds^{2}=-X^{2}dT^{2}+dX^{2}+dY^{2}+dZ^{2}.
\]
The observer $O_{a}$ is ``at rest'' at the space point $(X,Y,Z)=(1/a,0,0)$.
The space coordinates $(X,Y,Z)$ has the usual Euclidean metric, which
is invariant w.r.t.~the time translation $T\mapsto T+t_{0}$, That
is, this time translation is a \termi{Born rigid motion} on $\cR$
(sometimes simply called a \termi{rigid motion} \cite{Giu2010}).
Moreover this coordinate is rotation-free. Thus the Rindler coordinate
system is considered to be a ``natural coordinate system for $O_{a}$''
in the above sense.\footnote{In considering the ``natural coordinate system'' for an observer,
also notice the notion of \termi{Fermi--Walker transport} \cite{Ryd2009}.}

Of course, this argument is not entirely robust yet; Besides the rigid
laboratory is not physically realistic as mentioned above, it should
be demonstrated that it is natural for the observer $O_{a}$ to consider
only the Rindler wedge $\cR$, not the whole Minkowski spacetime $\cM_{4}$.\footnote{Another questionable point is as follows. The location of the clock
within the laboratory causes slight variations in the time scale,
which affects the description of observational results, because the
proper time of the clock depends on its $X$-coordinate. This situation
does not seem very ``natural'' from our everyday perspective.

This would not be an issue if we idealized the laboratory size to
be \textquotedblleft infinitely small\textquotedblright{} (effectively
a single point). However, quantum fields are thought to only make
sense in extended regions in spacetime, so probably this will be an
over-idealization. For the time being, we shall proceed assuming the
laboratory is not excessively large.}

The hyperplanes $t=\pm x$ divide the Minkowski space $\cM_{4}$ into
the four parts: the right Rindler wedge $\cR=\cR_{{\rm R}}$ and
\[
\cR_{{\rm L}}:=\{(t,x,y,z)\in\R^{4}:\ x<-|t|\},
\]
\[
\cB:=\{(t,x,y,z)\in\R^{4}:\ t>|x|\},
\]
\[
\cW:=\{(t,x,y,z)\in\R^{4}:\ t<-|x|\}.
\]
$\cR_{{\rm L}}$ is called the \termi{left Rindler wedge}. The region
$\cR_{{\rm L}}$ is always spacelike (causally) separated from $O_{a}$,
and hence indeed it will be natural for $O_{a}$ to ignore $\cR_{{\rm L}}$.

The observer $O_{a}$ in the right Rindler wedge $\cR$ can send signals
to another observer in the region $\cB$, but $O_{a}$ will never
receive a reply. Therefore, for $O_{a}$, the region $\cB$ is like
the inner side of a black hole's \textquotedblleft event horizon\textquotedblright{}
and every event in $\cB$ remains unknowable for $O_{a}$. Let us
call $\cB$ the \termi{black-hole-like region} for $O_{a}$. Conversely,
an observer $O'$ in $\cW$ can send signals to $O_{a}$ in $\cR$,
but $O'$ will never receive a reply. Let us call $\cW$ the \termi{white-hole-like region}
for $O_{a}$. Thus, roughly speaking, all of three regions $\cR_{{\rm L}},\cB,\cW$
are not accessible from $O_{a}$, and so it is fairly natural for
$O_{a}$ to ignore them; The right Rindler wedge $\cR$ is the only
``accessible world'' for $O_{a}$.

However note that $O_{a}$ can look inside the white-hole-like region
$\cW$; e.g., $O_{a}$ can see the stars in $\cW$. If the proper
acceleration $a$ of $O_{a}$ is over $10\,{\rm m}\cdot{\rm s}^{-2}$,
a value close to the gravitational acceleration on Earth, the distance
between $O_{a}$ and the ``horizon'' is approximately less than
one light-year. Hence actually most stars that $O_{a}$ sees will
be in $\cW$. Thus it feels somewhat strange to say that $O_{a}$
can/should ignore all the objects and the events in $\cW$.

What is more perplexing is the following. We assumed that $O_{a}$
accelerates eternally. However this assumption is unusual in that
this contains the assumption on the future of $O_{a}$. Furthermore,
$O_{a}$'s future will depend on $O_{a}$'s free will; that is, it
is natural to think that $O_{a}$ can stop accelerating at will. We
can also consider the cases where the the acceleration of $O_{a}$
stops accidentally, e.g., by a rocket engine trouble, independent
of $O_{a}$'s will. In any case, if $O_{a}$ stops accelerating, $O_{a}$
will go into the region $\cB$, and so the interior of $\cB$ is no
longer ``unknowable'' for $O_{a}$. I previously stated that it
is fairly natural for $O_{a}$ to ignore the region $\cB$, but now
we find that this assertion is based on the unnatural assumption that
$O_{a}$ accelerates eternally in the future.

Probably the unnaturalness of this type will be inevitable if we describe
thermodynamic notions in terms of global KMS states in QFTCS (or QFTM),
which is defined on the whole spacetime, including the far future
and the far past; Here we cannot speak about the temperature of a
local physical object, but about the ``temperature of the whole (past/present/future)
Universe''. This is quite different from our everyday understanding
of the term ``temperature''. Notice that Buchholz, Ojima and Roos
\cite{BOR2002} proposed a notion of \emph{local} temperature in QFTM.
However, a discussion of this proposal will be reserved for another
occasion.

Although currently I cannot find a good resolution for the above conceptual
difficulties, tentatively I will agree that the Rindler coordinate
system is a relatively natural reference frame for a uniformly accelerating
observer. However, usually the Unruh effect is formulated with the
Rindler coordinate system as in Claim \ref{claim:Unruh-xi}, and hence
the conceptual basis of the Unruh effect will suffer from these problematic
points.

\section{The Bisognano\textendash Wichmann theorem and the Unruh effect}

\subsection{Axioms for QFTM}

Recall some axiomatic foundations of QFTM.

The \termi{G{\aa}rding--Wightman axioms} (or the \termi{Wightman axioms})
for QFT in the Minkowski space $\cM_{4}\cong\R^{4}$ are roughly stated
as follows. \cite{SW64}.
\begin{ax}
{}
\begin{enumerate}
\item %
{} (a) A (pure) state is represented by a unit vector in a Hilbert space
${\mathcal{H}}$. (b) The Poincar\'e group acts on $\cH$ (i.e. we
are given a unitary representation of the Poincar\'e group on $\cH$).
\item %
{} The 4-momentum $(P_{0},P_{1},P_{2},P_{3})$ is positive-definite,
i.e., its spectrum is contained in the forward light cone. (spectrum
condition).
\item %
{} There exists a unique Poincar\'e invariant state $\Omega\in\cH$
(up to scalar multiple). (vacuum).
\item %
{} A quantum field $\phi$ is an operator-valued distribution defined
on a dense domain $\cD\subset\cH$.
\item %
{} $\phi$ transforms covariantly w.r.t.~the Poincar\'e transformations.
\item %
{} Two spacelike separated field operators are either commutative or
anticommutative.
\end{enumerate}
\end{ax}

On the other hand, the \termi{Araki--Haag--Kastler (AHK) axioms}
\cite{Haa96,Araki99} are roughly and intuitively stated as follows. 
Consider a description of a QFTM on a Hilbert $\cH$. (Here we use
von Neumann algebras on $\cH$, rather than abstract $C^{*}$-algebras.)
Let $\SYM{B(\cH)}{B(H)}$ denote the {*}-algebra of all bounded operators
on $\cH$. For an open set $D\subset\cM_{4}$, let $\SYM{\scO(D)}{O(D)}\,(\subset B(\cH))$
be the von Neumann algebra generated by the field operators whose
support is in $D$. Let $\SYM{\fA}A$ be the $C^{*}$-algebra generated
by $\bigcup_{D}\scO(D)$, where $D$ runs through all bounded open
subsets of $\cM_{4}$. 

Let $\scP_{+}^{\uparrow}$ denote the Poincare group (=inhomogeneous
Lorentz group). The transformation w.r.t.~$g\in\scP_{+}^{\uparrow}$
is represented by an automorphism $\alpha_{g}:\fA\to\fA$ on $\fA$.
\begin{ax}
[local observables]
\begin{enumerate}
\item uniformity: $D_{1}\supset D_{2}\then\scO(D_{1})\supset\scO(D_{2})$
\item covariance: $\forall g\in\scP_{+}^{\uparrow}$, $\alpha_{g}[\scO(D)]=\scO(g(D))$
\item locality: if $D_{1},D_{2}$ are spacelike separated, $\scO(D_{1})$
and $\scO(D_{2})$ are commutative.
\end{enumerate}
\end{ax}

Roughly speaking, the Wightman axioms are stronger than the Araki\textendash Haag\textendash Kastler
axioms; in other words, the latter is more general than the former.
However, strictly speaking, this is not the case; If we add some technical
assumptions on the Wightman axioms, then the AHK axioms follows from
them. Let us call such stronger Wightman axioms the \termi{Wightman--AHK axioms}. 

\subsection{The Bisognano\textendash Wichmann theorem}

Let $\cR$ be the (right) Rindler wedge. Let $\omega_{\cR}$ be the
state on $\scO(\cR)$, defined to be the restriction of the vacuum
$\Omega\in\cH$ onto $\cR$. For $r\in\R$, let $\Lambda(r)$ be the
Lorentz boost on $\cM_{4}$ defined by
\[
\Lambda(r):=\begin{pmatrix}\cosh r & \sinh r & 0 & 0\\
\sinh r & \cosh r & 0 & 0\\
0 & 0 & 1 & 0\\
0 & 0 & 0 & 1
\end{pmatrix}.
\]

\begin{thm}
[Bisognano--Wichmann theorem (outline) \cite{BW1975}, Theorem 4.1.1
in \cite{Haa96}]\label{thm:BW} Assume that $D\mapsto\scO(D)$ satisfies
the Wightman\textendash AHK axioms. Then $\omega_{\cR}$ is a $(\sigma,1)$-KMS
state on $\scO(\cR)$ for a unique group of modular automorphisms
$\sigma_{s}$ ($s\in\R$) on $\scO(\cR)$. Moreover $\sigma_{s}$
is nothing other than the action of the Lorentz boost $\Lambda(2\pi s)$
on $\scO(\cR)$. 
\end{thm}

Note that
\[
\Lambda(2\pi s)\begin{pmatrix}0\\
x\\
0\\
0
\end{pmatrix}=\begin{pmatrix}x\sinh2\pi s\\
x\cosh2\pi s\\
0\\
0
\end{pmatrix},
\]
and that the Rindler coordinates (\ref{eq:RindlerCoord}) is expressed
as
\[
\begin{pmatrix}t\\
x\\
0\\
0
\end{pmatrix}=\Lambda(aT)\begin{pmatrix}0\\
X\\
0\\
0
\end{pmatrix}=\begin{pmatrix}X\sinh aT\\
X\cosh aT\\
0\\
0
\end{pmatrix}.
\]
Thus the Lorentz boost $\Lambda(2\pi s)$ is equal to the time translation
$T\mapsto T+(2\pi/a)s$ w.r.t.~the Rindler coordinates. The ``speed
of the modular time evolution'' $ds/dT$ relative to the time parameter
$T$ equals to $a/(2\pi)$. Recall that $T$ is the ``natural time
scale'' for the observer $O_{a}$. Since the temperature relative
to the time scale $s$ is $\Temp=1$, the temperature relative to
the time scale $T$ is $\Temp_{{\rm U}}=a/(2\pi)$ (recall Claim \ref{claim:temp=00003Dtime}).
Thus it is natural to interpret that the ``temperature of the vacuum
universe'' viewed from the uniformly accelerating observer $O_{a}$
is equal to the Unruh temperature $\Temp_{{\rm U}}$, although this
``naturalness'' is somewhat doubtful as mentioned in Sec.~\ref{sec:Rindler-coordinate}. 

Bisognano and Wichmann \cite{BW1975} appeared in 1975, earlier than
Unruh \cite{Unr1976}, but they gave no such thermal interpretation
of this theorem. The Bisognano\textendash Wichmann theorem was reformulated
as a statement of the Unruh effect, similar to Claim \ref{claim:Unruh-xi},
by Sewell \cite{Sew1982} in 1982. Earman \cite{Ear2011} calls this
formulation the \termi{modular theory approach} to the Unruh effect.
Whereas Earman \cite{Ear2011} basically supports the modular theory
approach, he also acknowledges some serious conceptual problems on
this approach \cite[Sec.5]{Ear2011}.

Unruh himself \cite{Unr1976} and most of his followers have examined
the Unruh effect in significantly simplified, rather unrealistic models,
typically, a single accelerating particle interacting with a free
massless scalar field. On the other hand, the modular theory approach
has the advantage of being model-independent. However, it seems unlikely
that the results on such specific models, including those of \cite{Unr1976},
can be immediately derived from the modular theory approach. Hence
we cannot simply say that the modular approach is more general than
other approaches to the Unruh effect. Also note that the Wightman
axioms for QFTM is not quite general in that they do not cover gauge
field theories (including free electromagnetic fields). See \cite[Appendix]{Hor1990}.
It is expected that the Bisognano\textendash Wichmann theorem holds
also for a large class of non-Wightman fields in Minkowski spacetime
(c.f.~Remark on Theorem 4.1.1 in \cite{Haa96}). For partial results,
see Morinelli \cite{Mor2018}.

\subsection{Conceptual difficulties of the modular approach}

I already mentioned the conceptual difficulties of the Rindler spacetime
approaches to the Unruh effect in Section \ref{sec:Rindler-coordinate}.
The above modular approach, based on the Bisognano\textendash Wichmann
theorem \ref{thm:BW} together with the Tomita\textendash Takesaki
theorem \ref{thm:TT}, is one of the Rindler spacetime approaches,
and hence it suffers from those difficulties; This approach assumes
the eternal acceleration of the observer, but generally it is very
strange to make an assumption on the future, in physics. Usually,
the success or failure of a physical theory is judged by its predictive
power, that is, the ability to predict the unknown future from the
known data at the past and the present. Since we cannot know the far
future of an observer (or an object) in principle, one may doubt that
such an assumption on the future is logically meaningful. If we cannot
know the validity of an assumption, what is the meaning of the consequences
of the assumption?

\providecommand{\noopsort}[1]{}\providecommand{\singleletter}[1]{#1}%

\end{document}